\documentclass[useAMS,usenatbib]{mn2e}
\usepackage{graphicx}
\usepackage{aas_macros}

\title[Jet-dominated advective systems]{Jet-dominated advective systems: radio and X-ray luminosity dependence on the accretion rate}
\author[E. G. K\"ording, R. P. Fender and S. Migliari]{E. G. K\"ording$^{1}$\thanks{E-mail:
Elmar@phys.soton.ac.uk}, R. P. Fender$^{1}$ and S. Migliari$^{2}$\\
$^{1}$ School of Physics and Astronomy, University of Southampton Hampshire SO17 1BJ, United Kingdom\\
$^2$ Center for Astrophysics and Space Sciences, University of California San Diego, 9500 Gilman Dr., La Jolla, CA 92093-0424\\}
\begin{document}

\date{Accepted: 27.3.2006  }

\pagerange{\pageref{firstpage}--\pageref{lastpage}} \pubyear{2006}

\maketitle

\label{firstpage}

\begin{abstract}
We present a novel method to measure the accretion rate of radio emitting X-ray binaries (XRBs) and active galactic nuclei (AGN) independently of the X-ray luminosity. The radio emission of the jet is used as a tracer for the accretion rate and is normalised using sources of known accretion rates: island state neutron stars and efficiently radiating black holes close to a state transition. We show that the radio power in black holes and neutron stars is comparable for a given mass accretion rate and verify empirically the assumed analytic scaling of the radio luminosity with accretion rate ($L_{Rad}\propto \dot{M}^{1.4}$). 
As our accretion measure is independent of the X-ray luminosities, we can search for radiatively inefficient accretion in black holes by comparing the X-ray luminosities with the accretion rate in XRBs and AGN. While the X-ray luminosity of efficiently radiating objects scales linearly with accretion rate, the scaling of hard state black holes is quadratical, in agreement with theoretical models. We show that the turnover from the inefficient quadratic scaling to the linear scaling has to occur at accretion rates of 1-10\% Eddington both in XRBs and AGN. 
The comparison of both accretion states supports the idea that in a black hole in the hard state some accretion power is advected into the black hole while the jet power exceeds the X-ray luminosity: these are therefore jet-dominated advective systems.
\end{abstract}


\section{Introduction}
Measuring the accretion rate is one of the key issues in the research of accreting systems containing compact objects. 
Observationally, the accretion rate and the mass of the central object seem to be the most important fundamental parameters of the accreting system (see e.g., \citealt{BorosonGreen1992,UrryPadovani1995,MerloniHeinzdiMatteo2003,FalckeKoerdingMarkoff2004}). 

Black hole (BH) X-ray binaries (XRBs) can be found in a number of different accretion states. The two most prominent ones are the hard state, where the X-ray spectrum is dominated by a hard power-law  and the soft state, dominated by a soft thermal component (c.f., \citealt{Nowak1995,HomanBelloni2005,McClintockRemillard2003}).
In the soft state, the thermal component is usually interpreted as the signature of a standard thin disc \citep{ShakuraSunyaev1973}. As these discs are assumed to be radiatively efficient, one can estimate the accretion rate from the bolometric luminosity. In the hard state, however, the disc is likely to be radiatively inefficient (e.g., advection dominated accretion flows, ADAF: \citealt{NarayanYi1994,EsinMcClintockNarayan1997}) and the bolometric luminosity will underestimate the accretion rate. Some of the accretion power may cross the event horizon and cannot escape to infinity.  Besides the soft and the hard state intermediate 'transition' states, usually short-lived, are also found. These intermediate states (IMS) occur in two "flavors", a hard IMS characterised by a hard spectral component and band-limited noise in the power spectrum and a soft IMS dominated by a soft spectral component with power law noise in the power spectrum (e.g., \citealt{BelloniHomanCasella2005}). 

In neutron star (NS) binaries one can always estimate the accretion rate from the bolometric luminosity, as the accreted matter has to hit the stellar surface and the kinetic energy will be radiated, although it seems there is not a perfect one-to-one relation between the X-ray luminosity and other accretion rate proxies (see \citealt{Klis2001} and references therein). Similarly to the BHs, NSs also show different X-ray states. Low magnetic field NS systems are typically divided into atoll and Z-sources \citep{HasingerKlis1989}. Z-sources are strongly accreting (0.5 - 1 Eddington rate) while atoll sources probably have lower accretion rates. The analog of hard state for BHs is the island state of atoll NSs, while the soft state is attributed to the banana state for atoll NSs. This analogy is also supported by variability studies (see \citealt{Klis2004} for a review). The Z-sources are usually attributed to a high luminosity IMS.

While a source is in its hard state, one usually observes a flat spectrum radio core \citep{Fender2001}, which is a signature of a relativistic jet. The radio emission of NSs seems to be lower than that of a BH for a given X-ray luminosity \citep{MunoBelloniDhawan2005,MigliariFender2005b}. For BHs this radio emission is quenched once the source enters the high state \citep{TananbaumGurskyKellogg1972,FenderCorbelTzioumis1999}. NSs in the banana state, however, do show radio emission (e.g., \citealt{MigliariFenderRupen2004}). The radio emission in the hard state seems to be a good tracer for the accretion rate \citep{FenderGalloJonker2003}. It has already been shown that the radio emission can be used to estimate the jet power of a source \citep{FalckeBiermann1999,HeinzGrimm2005}. We will derive a method to use the radio luminosity as a measure for the accretion rate and normalise it with sources of known accretion rate. 

The derived accretion rates depend only on the radio luminosities and are, except for the normalisation points, independent of the X-ray fluxes. Thus, we can use this accretion rate to compare the X-ray emission from NSs and BHs and search for hints of radiatively inefficient accretion in BHs, which may be a signature of a BH event horizon (see \citealt{NarayanGarciaMcClintock1997}, but note \citealt{AbramowiczLasota2002}).

\section{Radio as a tracer of the accretion rate}

\subsection{Deriving the accretion rate}\label{seaccretion}
\subsubsection*{Soft state} 
Accreting objects in their soft state show a strong multi-temperature black body component in their spectral energy density (SED), usually explained by an optically thick, efficiently radiating standard disc \citep{ShakuraSunyaev1973,McClintockRemillard2003}. Assuming a disc efficiency $\eta$, we can derive the accretion rate from the luminosity:
\begin{equation}
\dot{M} =  \frac{L_{\mathrm bol}}{f \eta c^2} = 1.5 \times 10^{18} \left(\frac{L_{\mathrm bol}}{10^{38}\mbox{erg s}^{-1}} \right)\left(\frac{0.75}{f}\right)\left(\frac{0.1}{\eta}\right) \frac{\mbox{g}}{\mbox{s}}
\label{eqsoftdisc},
\end{equation}
where $f$ reflects what part of the total accretion rate external to the radiating region is actually accreted and is not ejected in the jet or the winds. The factor $f$ can depend on the accretion rate; we assume that $f$ is similar for our objects at a given $\dot{M}$. Throughout this paper we will use $\eta = 0.1$ and $f = 0.75$, but we will give the final accretion rate measures also for arbitrary values of $\eta$ and $f$. 

\subsubsection*{Hard state}
The method used to calculate the accretion rate for soft state objects can not be used for the hard state, as the accretion flow is likely to be radiatively inefficient. Here, the radio luminosity can be used as a tracer of the accretion rate (e.g., \citealt{FenderGalloJonker2003}). A simple conical jet model (see e.g., \citealt{BlandfordKonigl1979,FalckeBiermann1995}) yields for the optically thick radio emission:
\begin{equation}
L_{\mathrm Rad} \propto P_{\mathrm j}^{17/12}, 
\end{equation}
where $L_{\mathrm Rad} = \nu F_\nu$ for the observing frequency $\nu$. Jets in hard state XRBs do not seem to be highly relativistic, thus relativistic beaming is of minor importance (\citealt{GalloFenderPooley2003}, but see \citealt{HeinzMerloni2004}).
We assume that the accretion disc and the jet form a symbiotic system, so that the jet power is always a constant fraction of the accretion rate $P_{\mathrm j} = q_{\mathrm j} \dot{M}$ \citep{FalckeBiermann1995}. This assumption will be tested later. Thus, we find:
\begin{equation}
\dot{M} = \dot{M}_{\mathrm 0} \left(\frac{L_{\mathrm Rad}}{L_{\mathrm Rad, 0}} \right)^{12/17}, \label{eqMdotrad}
\end{equation}
where the normalisation factors ($\dot{M}_{\mathrm 0}$, $L_{\mathrm Rad, 0}$) have to be determined. As both constants are exchangeable we set $L_{\mathrm 8.6 GHz, 0} = 10^{30}$ erg s$^{-1}$. This is roughly the 8.6 GHz radio luminosity where the accretion disc around a $10 M_\odot$ BH changes its spectral state.

\subsection{The sample}
To find the normalisation $\dot{M}_0$ of the relation (\ref{eqMdotrad}), we use a sample of radio-loud objects, with well measured accretion rates.
\subsubsection*{Hard state black holes}
The X-ray luminosity of hard state BHs probably does not originate from an efficient accretion flow, so we can not use an arbitrary X-ray measurement to obtain the accretion rate. However, we can use the fact that the total luminosity, and therefore also the accretion rate, does not strongly change during the state transition \citep{ZhangCuiHarmon1997,BelloniParolinSanto2006}. Once the source is in the soft state, we can estimate the accretion rate from the bolometric luminosity (eq. \ref{eqsoftdisc}). 
Thus, we will use a measurement of the radio and X-ray flux just before the source changes its state (the brightest hard state measurement). As we know that the bolometric luminosity does not change significantly, we can use the X-ray luminosity to obtain the accretion rate. The key assumption here is that $\dot{M}$ changes smoothly and relatively slowly across the state transition.

During the state transition from the hard to the soft state the 1-200 keV luminosity of Cyg X-1 remains unchanged within 15 \% \citep{ZhangCuiHarmon1997}. This luminosity is roughly 2 \% Eddington \citep{diSalvoDoneBurderi2001}. 
%This value appears to be universal for stellar black holes \citep{Maccarone2003}. 
The 5 GHz radio flux is around 15 mJy (e.g., \citealt{GleissnerWilmsPooley2004}) and the distance is 2.1 kpc \citep{MasseyJohnsonDegioiaEastwood1995}. 

\citet{CorbelNowakFender2003} found GX~339-4 in 1997 in a very bright hard state. It had a 8.6 GHz radio flux of 9.1 mJy. The 3-200 keV flux is $6.1 \times 10^{-9}$ erg s$^{-1}$ cm$^{-2}$ \citep{NowakWilmsHeinz2005}. \citet{GalloFenderPooley2003} found that the data-point coincides with the point where the radio emission starts to be quenched. As the source is in the hard state (photon index $\approx 1.6$), the bolometric luminosity will not be considerably higher than 3-200 keV luminosity. 
\citet{BelloniHomanCasella2005} followed GX~339-4 through a full cycle of state transitions during the 2002-3 outburst. On MJD52382.8 \citet{GalloCorbelFender2004} observed the source in radio and found a 8.6 GHz flux of 13.5 mJy. On that date the source was still in the hard state and had a PCU count rate of $\approx 890$ counts s$^{-1}$. The X-ray flux is only 11\% below the maximal flux before the state transition. This corresponds to a total X-ray flux of $2.8 \times 10^{-8}$ erg s$^{-1}$ cm$^{-2}$ using a bolometric correction of 1.34 for the 2-60 keV fluxes (the bolometric correction has been obtained from \citealt{CorbelNowakFender2003}). We use both data-points for GX~339-4. 
The distance to GX~339-4 is not well constrained; \citet{HynesSteeghsCasares2004} gives a lower limit of 6 kpc, but the distance may be as high as 15 kpc. We adopt a distance of $8$ kpc. 

Besides the three well constrained data-points just described we also add two additional points where the time of the state transition is less certain or have larger observational uncertainties.
Radio observations of V404 Cygni in its 1989 outburst first showed a steep radio spectrum, that turned into a flat spectrum on MJD 47680 \citet{HanHjellming1992}. If we associate a state transition back to the hard state with this change we can use the radio and X-ray observation just after this change. We use the radio flux reported by \citet{HanHjellming1992} and the X-ray flux of \citet{GalloCorbelFender2004} and assume a bolometric correction of 5 for the 2-10 keV flux. We adopt a distance of 9 kpc \citep{JonkerNelemans2004}. \citet{BrocksoppFenderMcCollough2002} observed XTE J1859+226 during a state transition from the hard to the soft state in radio and X-rays. Here we use a distance of 6.3 kpc \citep{JonkerNelemans2004}. We include these additional points with the same weight as the well constrained points in our fits. The inclusion of these less constrained objects does not significantly change the fitted normalisation constant (changes below 2\%).

For the hard state BHs we set the relative uncertainty of the radio and X-ray fluxes to 30\%, in order to reflect possible changes in the accretion rate during the transition and that we may not have the highest possible X-ray and radio fluxes before the transition.

\subsubsection*{Intermediate state black holes}
GRS~1915+105 in its radio plateau state shows a stable optically thick radio jet (e.g., \citealt{FenderGarringtonMcKay1999}). Even though the X-rays may arise from an inefficient flow, the observed X-ray flux is near the Eddington limit. As the source is always around the state transition and we know that the X-ray luminosities do not change significantly during a state transition, we will attempt to use the X-ray flux to derive the accretion rate. We use the radio and X-ray data during the plateau state as published by \citet{MunoRemillardMorgan2001} and assume that the soft power law in the X-ray spectrum has a lower cut-off at 1 keV. To get a single data-point, we average the data in log-space.  We assume the distance to be 11 kpc ( \citealt{FenderGarringtonMcKay1999,DhawanMirabelRodrguez2000}, but see also \citealt{ChapuisCorbel2004,KaiserGunnBrocksopp2004} for lower values).

\subsubsection*{Neutron Stars}
As NSs have a stellar surface, all accretion energy has to be radiated or carried away by winds or the jet. Assuming that the majority of the energy release is radiation, we can use all NSs in an analog state of the hard state, i.e., the island state, to normalise our accretion rate measure (eq. \ref{eqMdotrad}). We use all island state sources from \citet{MigliariFender2005b}. This sample is mainly based on the NS 4U~1728-34. We use a bolometric correction for the 2-10keV fluxes of a factor of 2.5 \citep{MigliariFender2005b}.

\begin{figure}
\resizebox{8.7cm}{!}{\includegraphics{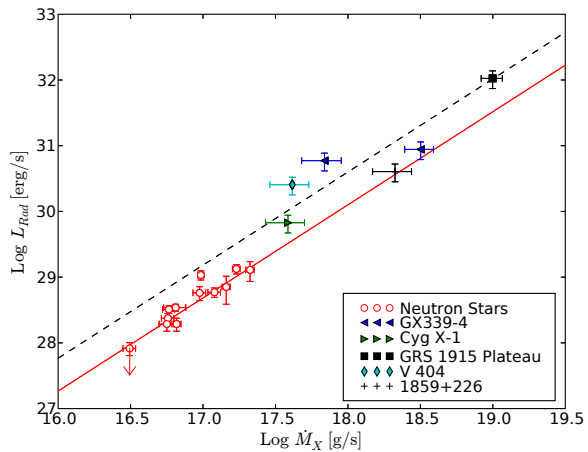}}
\caption{Radio luminosity as a function of the accretion rate measured with the bolometric X-ray luminosity. For hard state BHs we use measurements near the state transition to avoid the effect of the inefficient accretion flow, see text. The solid line denotes the analytical jet model fitted to the NSs, assuming that a constant fraction of the accretion power is injected into the jet. The radio varies in that case as $\dot{M}^{17/12}$. The dashed line shows model fitted to the BHs. }
\label{FigLXLrad}
\end{figure}

\subsection{The resulting accretion measure}
In Fig.~\ref{FigLXLrad} we show the radio luminosity as a function of the accretion rate as measured by the bolometric luminosities. 
The solid line represents the analytical jet model (i.e., $L_{\mathrm Rad} \propto \dot{M}^{17/12}$) fitted to the NSs, while the dashed line is fitted to the BH points. The analytical model assumes that a constant fraction of the accretion power is always injected into the jet. The model described in eq.~(\ref{eqMdotrad}) can be translated to
$\log L_{\mathrm Rad} = \frac{17}{12} \log \dot{M} + b_{RX}$, 
where the radio luminosity is measured in erg s$^{-1}$ and the accretion rate in g s$^{-1}$ and $b_{RX} = \left(\log L_{Rad,0}/\dot{M}_0^{17/12}\right)$. For the NSs we find $b_{RX}^{NS} = 4.7 \pm 0.13$ while the BHs give $b_{RX}^{BH} =  5.1\pm 0.29$. The uncertainty given is the sample standard deviation, which is of minor significance for the BHs as we only have 4 points. As the NS sample is dominated by one source (4U 1728-34), the scatter around the correlation is not a good measure of the actual scatter of the normalisation constant and therefore of $M_0$. We fixed the slope of the relation to the theoretical prediction, for NSs \citet{MigliariFender2005b} fitted the dependence of the radio luminosity on the X-ray luminosity in the island state sources , and found that $L_{Rad} \propto L_X^{1.4}$ which is exactly the relation found analytically for conical jets. 

The hard state BHs are scattering around the correlation. This scatter is not only due to measurement or distance errors, as the two points for GX339-4 seem to follow a flatter slope than found in NSs. 
If a hard state BH is inefficiently radiating, the point will deviate from the correlation in Fig.~\ref{FigLXLrad} towards lower X-ray luminosities. Inefficiently radiating BHs scale with $L_{Rad} \propto L_{X}^{0.7}$ \citep{CorbelNowakFender2003,GalloFenderPooley2003} compared to $L_{Rad} \propto L_X^{1.4}$ found for efficiently radiating objects. Thus, if a source is not observed exactly at the hard to soft state transition but shortly before (or in case of the soft to hard transition after the transition), the point will lie above the predicted correlation. 

In BH XRBs a hysteresis effect has been observed: the transition luminosity from the hard to the soft state is higher than that of the transition back \citep{Maccarone2003,HomanBelloni2005}. Also the hard to soft state transition is not at the same Eddington ratio in each outburst \citep{BelloniParolinSanto2006}. 
If the source is always radiatively efficient in its soft state, this has to lead to parallel track in the radio/X-ray correlation if the jet coupling remains fairly constant. There is some indication of these parallel tracks in GX339-4 \citep{NowakWilmsHeinz2005}. These parallel tracks would lead to horizontal scatter in the radio/X-ray correlation of efficiently radiating objects (Fig.~\ref{FigLXLrad}). Our two data-points for GX339-4 are from two different outbursts and have been used to argue for parallel tracks.  
Summarised, the scatter around the correlation is likely to be partly due to still inefficiently radiating objects and due to the hysteresis effect. 

The BHs in Fig.~\ref{FigLXLrad} lie above the fitted line for NSs, by 0.4 dex or a factor of 2.5. This excess by a factor 2.5 found for the radio luminosity for BHs corresponds to a factor 2 in jet power. The real difference between NSs and BHs may be even lower, if we did not use the highest possible radio and X-ray fluxes in our sample.
It is unlikely that the normalisations of NSs and BHs are exactly the same, because the inner parts of the accretion flow have to be different, e.g., NSs should have a boundary layer and NS accretion discs will be truncated at larger radii than those of spinning BHs. However, we can not be sure that we measured this as our errors are just a lower limit to the real uncertainty, so the difference is not significant.

The radio/accretion rate correlation found for NSs, hard state BHs and IMS BHs  showing an optically thick radio jet translates to normalisation factors $\dot{M}_0$ for our method to derive the accretion rate from the radio luminosity (\ref{eqMdotrad}): 
\begin{equation}
\dot{M}^{NS}_0 = 7.7 \times 10^{17} \mbox{g s}^{-1} \qquad  \dot{M}^{BH}_0 = 4.0 \times 10^{17} \mbox{g s}^{-1}.
\end{equation}
Eq.~(\ref{eqMdotrad}) together with this normalisation allows us to estimate the accretion rate from the radio luminosity.

\subsection{Comparison to measures of the jet power}
We have found a measure of the accretion rate using the radio luminosity of the optically thick jet, which can be compared with the jet power estimates from the radio luminosity. The jet power at the state transition can be at most in equipartition with the bolometric luminosity, as otherwise one would observe a jump in the radiative luminosity during the state transition to the soft state. Thus, we find for black holes 
\begin{eqnarray}
P_{jet} &\leq& f \dot{M}_{\mathrm 0}^{BH} \eta c^2 \left(\frac{L_{8.6 GHz}}{L_0}\right)^{12/17} \\ 
&\approx& 3.6 \times 10^{37} \left(\frac{f}{0.75}\right) \left(\frac{\eta}{0.1}\right) \left(\frac{L_{8.6 GHz}}{L_0}\right)^{12/17} \mbox{erg s}^{-1}. \nonumber
\end{eqnarray}

\citet{HeinzGrimm2005} calculate the total jet power from the radio luminosity. Their normalisation is based on XRBs and AGN, some of which are in the IMS. They find that the jet power $P_{\mathrm jet} = 3.2 \times 10^{37} \ \mbox{erg s}^{-1}\  (L_{8.6 GHz}/L_0)^{12/17}$ for both XRBs and AGN. The normalisation value is also roughly in agreement with other theoretical jet models, see e.g., \citet{MarkoffFalckeFender2001}. This fairly high jet normalisation suggests, that the jet power is indeed near equipartition with the bolometric luminosity at the state transition. This has been assumed in \citet{MigliariFender2005b} and proposed as "total equipartition jet" in \citet{FalckeBiermann1995}. 

\citet{GalloFenderKaiser2005} calculate the total jet power of Cyg~X-1 from the observation of a ring like structure that was probably inflated by the radio jet. Their estimate for the average jet power is $9\times 10^{35} \mbox{erg s}^{-1} \leq P_{jet} \leq 10^{37} \mbox{erg s}^{-1}$. This is a time-averaged value of the jet power during the time the source needed to inflate the structure and we have to assume that the accretion rate stayed roughly constant over that time. It translates to (assuming an average radio flux of 15 mJy): $P_{\mathrm jet} = 1.2 - 13 \times 10^{36} \ \mbox{erg s}^{-1}\  (L_{Rad}/L_{Rad,0})^{12/17}$. This is well in agreement with our upper limit for the jet power. Combined with the result by \citet{HeinzGrimm2005} it is likely that the jet power is around the upper limit of the given range. This suggests:
\begin{equation}
P_{\mathrm jet} \approx \frac{1}{2} \eta f \dot{M} c^2,
\end{equation}
i.e., at all $\dot{M}$ about half of the liberated accretion power goes into the jet.
Under the assumption that the radiative efficiency in the radio band of BHs and NSs is similar, the proportionality factor for NSs will be slightly lower; our findings suggest $1/4$.

\section{Bolometric luminosity dependence on the accretion rate}
We have seen in Sect.~\ref{seaccretion} that we can use the radio luminosity to estimate the accretion rate. This measure seems to be applicable to every object with a steady, optically thick jet, even though the existence of optically thick emission in NSs is not yet proven as the uncertainties on the spectral indexes are still too large \citep{MigliariFenderRupen2003}.  As the estimate is independent of the measured X-ray luminosity we can now compare the X-ray flux of radio-loud objects with the accretion rate.

Hard state BH XRBs show a steady optically thick jet, so our method is applicable. Our hard state BH XRB sample is based on \citet{GalloFenderPooley2003} and consist of GX~339-4, V404 Cyg, XTE J1118+480, and 4U 1543-47. For all sources we use the same bolometric correction, which was derived for GX~339-4 (5.7 for the 3-9 keV luminosity,\citealt{CorbelNowakFender2003}). We include only measurements in the hard state.

\begin{figure}
\resizebox{8.7cm}{!}{\includegraphics{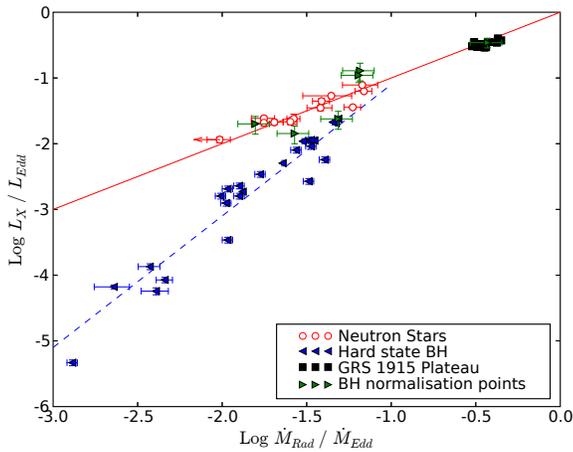}}
\caption{X-ray luminosity as a function of accretion rate for efficient and radiatively inefficient objects. The solid line denotes the linear dependence on the accretion rate for efficient discs, while the dashed line shows the quadratic dependence of the bolometric luminosity of inefficient flows or jets. }
\label{fiMdotRLX}
\end{figure}

Using eq. (\ref{eqMdotrad}) we calculate the accretion rate for our hard state BHs, GRS 1915+105 in the plateau state and the NS sample. In Fig.~\ref{fiMdotRLX} we show the bolometric X-ray luminosity divided by the Eddington luminosity as a function of the accretion rate in Eddington units. The X-ray luminosity of our NSs and  at least the radio loud  IMS objects depend linearly on the accretion rate, as expected from Sect.\ref{seaccretion}. The hard state BH XRBs, however, depend more strongly on the accretion rate and fall off roughly as $\dot{M}^2$. This is in agreement with models for the X-ray emission due to inefficient flows (see e.g., \citealt{Ichimaru1977,ReesPhinneyBegelman1982,NarayanYi1994}) and alternatively due to jet emission \citep{MarkoffNowakCorbel2003}.

NSs seem to be radiatively efficient, suggesting that the X-rays originate predominantly from the boundary layer and the accretion disc/corona system. The implications for the emission processes in hard state BHs are not straightforward and will be discussed in a future paper.  \citet{FenderKuulkers2001,MunoBelloniDhawan2005,MigliariFender2005b} showed that the 'radio loudness', i.e., $\frac{L_{\mathrm Rad}}{L_{\mathrm X}}$, is lower for NSs than for BHs. In the context of our findings, this is simply due to the lower X-ray luminosity of a hard state BH for a given accretion rate compared to the luminosity of a NS. 

\citet{GalloFenderPooley2003} and \citet{CorbelNowakFender2003} showed that the radio/X-ray correlation can be written as $L_X\propto L_{Rad}^{1.4}$. The radio luminosity scales with the accretion rate as $L_{Rad} \propto \dot{M}^{17/12}$; and we find:
$L_X\propto \dot{M}^{2}$. Thus, the slope of the hard state BHs in Fig.~\ref{fiMdotRLX} follows from the radio/X-ray correlation for BHs \citep{GalloFenderPooley2003} while the slope of the NSs is governed by the radio/X-ray correlation for NSs \citep{MigliariFender2005b}. As these papers already published detailed fits to the slopes, we will not present fits for the different source classes.

\subsection{Active Galactic Nuclei}
We have compared the X-ray luminosity of stellar objects with the accretion rates obtained from the radio luminosities and found a different scaling for efficient and inefficiently accreting objects. The X-ray luminosities we used for the XRBs in Fig.~\ref{fiMdotRLX} are bolometric luminosities. It is hard to obtain bolometric corrections for AGN as their SEDs are complex and the measured X-ray flux does not contain the majority of the total flux for many sources.  
Furthermore, the bolometric correction will probably depend on the BH mass. For jet models this can be seen in the spectral correction for different BH masses described in \citet{FalckeKoerdingMarkoff2004}; in Comptonization models the seed photon energy a will depend on the black hole mass and the optical depth and electron temperature may change as well.

Nevertheless, as a first approach we use the same bolometric correction as we use for hard state XRBs
 and compare AGN in the analog of the hard state (ie. below $\approx 1$ \% Eddington) to our stellar sample. The missing mass dependence will be introduced later. Thus, we will first assume that the bolometric luminosity is simply 5.7 times the 3-9 keV luminosity.  To measure the accretion rates, we use the same normalisation values as found for the Galactic sources. 

\begin{figure}
\resizebox{8.7cm}{!}{\includegraphics{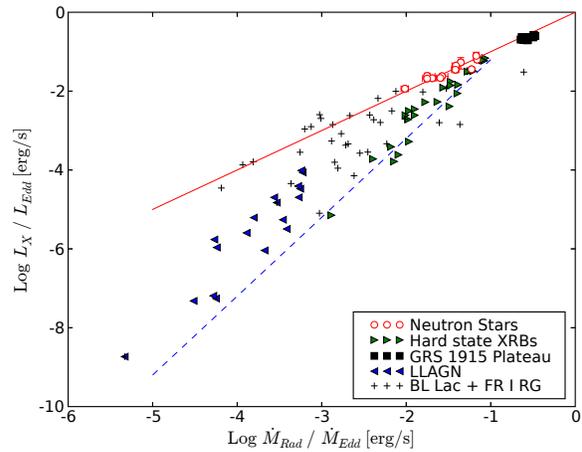}}
\caption{X-ray luminosity as a function of accretion rate for stellar and supermassive sources. We use the same bolometric correction for hard state XRBs and our hard state AGN sample. Only the LLAGN are all clearly below the linear dependence of efficient accretion.}
\label{FigUncorrAGN}
\end{figure}

We use the AGN sample of \citet{KoerdingFalckeCorbel2005}, which is based on the sample in \citet{FalckeKoerdingMarkoff2004}. Compared to the old sample the new sample contains all low ionisation nuclear emission region objects in the radio sample of \citet{NagarFalckeWilson2005}, where we found X-ray luminosities in the literature.
It contains low luminosity AGN (LLAGN), the Galactic center Sgr A$^*$, FR-I radio galaxies \citep{FanaroffRiley1974} and BL Lac objects.

In Fig.~\ref{FigUncorrAGN} we show this AGN sample together with our stellar sample using a constant bolometric correction for both XRBs and AGN. The LLAGN are all below the linear relationship of the efficiently radiating objects suggesting that they are inefficiently radiating like hard state XRBs. However, they all lie above the quadratic dependence of the hard state XRBs. The BL Lac and FR-I RGs are even at higher X-ray luminosities and some seem to follow the linear correlation for efficiently radiating objects, which is unlikely, as they are thought to be in the hard state (e.g., \citealt{FalckeKoerdingMarkoff2004}).

\begin{figure}
\resizebox{8.7cm}{!}{\includegraphics{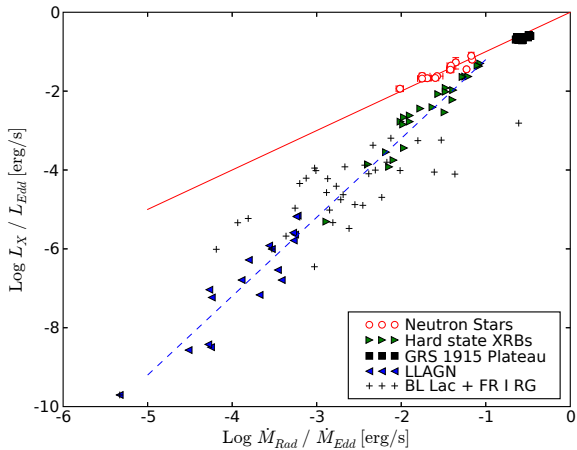}}
\caption{X-ray luminosity as a function of accretion rate for efficient and radiatively inefficient objects. Compared to Fig.~\protect\ref{FigUncorrAGN} we include the mass term of the fundamental plane of $M^{0.15}$. Now, all hard AGN follow the quadratic scaling of inefficient accretion flows or jets.}
\label{FigAGN}
\end{figure}

However, up to now, we have ignored that the bolometric correction of hard state black holes will depend on the BH mass. 
We can use the fundamental plane of accreting BHs \citep{MerloniHeinzdiMatteo2003,FalckeKoerdingMarkoff2004}, i.e., the extension of the radio/X-ray correlation for XRBs to the AGN, to create an 'effective' bolometric correction. The fundamental plane can be approximated by
$\log L_X = 1.4 \log L_{Rad} - 0.86 \log M + b$, where $b$ is the unimportant offset \citep{KoerdingFalckeCorbel2005}. This correlation can be rewritten as:
\begin{equation}
\frac{L_X}{L_{Edd}} \propto \left(\frac{\dot{M}}{\dot{M}_{Edd}}\right)^2 M^{0.14} . \label{eqFundamental}
\end{equation}
Thus, the fundamental plane differs only by a factor of $M^{0.14}$ from the simple expression for an inefficient accretion flow or a jet $\left(L_X/L_{Edd} \propto \left(\dot{M}/\dot{M}_{Edd}\right)^2\right)$ under the assumption that $L_{rad} \propto \dot{M}^{17/12} \sim \dot{M}^{1.4}$. 

Both models used to explain the fundamental plane (jet-only and ADAF/jet) can explain the mass term of $M^{0.86}$. They both use scale invariance as their fundamental assumption, thus the mass dependence of the bolometric luminosity has to be linear with $M$ (for a fixed Eddington ratio). The difference of $M^{0.14}$ in eq.~(\ref{eqFundamental}) is therefore an effect of the conversion of the observed X-ray luminosity to the bolometric luminosity. For a derivation of the X-ray and bolometric luminosities of a jet-only model see e.g., \citealt{Koerding2004}, a similar relation can probably be derived for inefficient accretion flow models. We therefore use the bolometric correction for hard state XRBs and add the additional mass term to obtain an 'effective' bolometric correction for our AGN sample. 

The resulting plot is shown in Fig.~\ref{FigAGN}. 
Now, the AGN sample and our hard state XRBs follow the same track (as shown in the papers on the fundamental plane, \citealt{MerloniHeinzdiMatteo2003,FalckeKoerdingMarkoff2004}). We have also fitted a power law to the data and find $\frac{L_X}{L_{Edd}} \propto  \left(\frac{\dot{M}}{\dot{M}_{Edd}}\right)^{2\pm 0.2} M^{0.15\pm 0.05}$, but the exact fit values depend on the assumptions mode on the distribution of the underlying scatter (see \citealt{KoerdingFalckeCorbel2005}).
This supports our approach that AGN and BH XRBs have roughly the same accretion rate normalisation $M_0$. One problem for AGN is that relativistic boosting seems to play a more important role in AGN than in hard state XRBs (e.g., the existence of Blazars). Our accretion rate measure needs intrinsic deboosted radio luminosities, and will therefore have larger uncertainties for AGN.

In the context of the ADAF model for the X-ray emission \citep{NarayanYi1994} we note that the mass term in the fundamental plane is -- except for the bolometric correction ($M^{0.15}$) -- a natural consequence of the scaling of the X-ray luminosity of inefficient flows with $(\dot{M}/\dot{M}_{Edd})^2 L_{Edd}$ and arises without requiring a spectral correction. In the jet only model, i.e. radio and X-rays are assumed to originate from the jet \citep{MarkoffFalckeFender2001}, the mass term can be explained by the spectral correction that connects the radio flux to the X-ray flux \citep{FalckeKoerdingMarkoff2004}. In fact this is analogous to the ADAF case, as the synchrotron X-ray emission of a jet scales roughly with $\approx (\dot{M}/\dot{M}_{Edd})^{1.8} M^{0.8}$ (the exact power law indexes depend on the parameters of the jet model).

\section{Discussion}
We have shown that our method to derive the accretion rate seems to work for island state NSs, hard state BHs and {radio loud} IMS BHs. In Fig.~\ref{FigLXLrad} we observe that the radio luminosity follows $L_{Rad} \propto \dot{M}^{17/12}$. As the radio luminosity scales with jet power as $L_{Rad} \propto P_{jet}^{17/12}$ \citep{BlandfordKonigl1979,FalckeBiermann1995}, it is therefore necessary that a similar fraction of the accretion power is always injected into the jet: $P_{jet} = q_j \dot{M} c^2$. The normalisation factor $q_j$ depends at most only slightly on the nature of the central object. For BHs it seems to be a factor of 2 larger than for NSs, this can either be due to a different efficiency but can also be understood as a signature of the boundary layer (which changes $\eta$). 
Thus, the jet does not depend strongly on the nature of the central object, but seems to be directly coupled to the accretion flow. The underlying jet-launching mechanism has to be nearly as efficient for NSs as for BHs. Any jet-launching mechanism for the hard state using properties only available in BHs (e.g., a strong BH spin dependence \citealt{BlandfordZnajek1977,LivioOgilviePringle1999,Meier2001}) or NSs (e.g., the surface dipole, magnetic fields) seem to be disfavoured by our findings (but note that there are NS versions of BH models and vice-versa).

The comparison between soft and IMS NSs and BH XRBs is more involved. There may be large differences between both classes, for a fuller comparison of NSs and BHs see \citet{MigliariFender2005b}, especially the discussion on Z-sources. 

Using the radio/X-ray correlation for BH XRBs one can translate our accretion rate measure using radio luminosities to one for the 2-10keV flux. We find for hard state black holes:
\begin{equation}
\dot{M}\approx 4.5 \times 10^{17} \left(\frac{L_{2-10keV}}{10^{36} \mbox{erg s}^{-1}} \right)^{0.5} \left(\frac{M}{M_{GX339}} \right)^{0.43} \frac{0.75}{f}\frac{0.1}{\eta} \frac{\mbox{g}}{\mbox{s}}
\end{equation}
As this relation is obtained by using the radio/X-ray correlation, it is likely that this measure has a higher uncertainty than the estimate using radio emission.
With this formula and eq.~(\ref{eqsoftdisc}) we can estimate the accretion rate for hard and soft state BHs from the X-ray flux.

In Fig.~\ref{fiMdotRLX}, we have seen that the bolometric luminosity of hard state BHs decreases with the accretion rate as $\dot{M}^2$, while the bolometric luminosity of NSs decreases linearly: 
\begin{equation}
L_{bol}^{NS} \propto \dot{M} \qquad L_{bol}^{BH} \propto \dot{M}^2
\end{equation}
It is tempting to attribute this difference to the existence of the event horizon, i.e., some of the accretion power can be transported into the BH, while for the NSs it always hits the stellar surface \citep{NarayanGarciaMcClintock1997}. To avoid having to scale everything to Eddington units in this discussion, let us assume that our BHs and NSs have roughly the same mass. 

The power created by accreting matter can either be radiated away, transported from the BH in the form of matter or magnetic fields (the jet and winds) or cross the event horizon. This can be written as:
\begin{equation}
\dot{M} \eta c^2 = P_{jet} + P_{wind} + L_{bol} + P_{Advect},
\end{equation}
where $\eta$ denotes the efficiency with that the BH can create energy from accretion (e.g., $\eta \approx 0.1$), $L_{bol}$ denotes the bolometric luminosity, $P_{jet}$ and $P_{wind}$ the power injected into the jet and wind and $P_{Advect}$ is the power advected into the BH. We attribute all matter and magnetic fields not contained in the jet to the accretion disc wind. Assuming that the radiative efficiency of BH and NS jets in the radio band is similar, we have seen that
\begin{equation}
P_{jet}^{NS}(\dot{M}) \approx P_{jet}^{BH}(\dot{M}) \approx q_j \dot{M},
\end{equation}
where the approximate sign $(\approx)$ means equal within a factor of a few.
Winds play an important role in strongly accreting objects (see e.g., \citealt{KingPounds2003}), but at least for radiation driven winds their role is smaller in our strongly sub-Eddington hard state BHs (e.g., \citealt{Proga1999}). Another possibility would be that the accretion disc is similar to the 'ADIOS' solution \citep{BlandfordBegelman1999}. Already in the original paper, these authors suggested as an observational test to check if NSs have inefficient flows, as the existence of inefficient NSs would strongly support their model. Our findings suggest that NSs are radiatively efficient and make it unlikely that the accretion flows in the hard state are of ADIOS type.
It is therefore likely, that either winds do not contribute significantly in the hard state or that similar to the jet production, the wind creation is a feature of the accretion flow and does not depend strongly on the central object, i.e., 
\begin{equation}
P_{wind}^{NS}(\dot{M})\approx P_{wind}^{BH}(\dot{M}).
\end{equation}
However, while this is an assumption that is likely to be fulfilled, we have not proven this. Combining these formula, we find:
\begin{equation}
P_{Advect}^{BH} \approx L^{NS}_{bol}(\dot{M}) - L^{BH}_{bol}(\dot{M})
\end{equation}
The difference between the linear dependence from the NSs and the quadratic dependence from the BHs in Fig.~\ref{fiMdotRLX} is likely to be advected into the BH -- under the assumption that the winds are similar (up to a factor of a few) in both systems.
In the case that the X-rays originate from the jet \citep{MarkoffFalckeFender2001} the advected fraction is even larger.
Thus, in the hard state black holes the power advected into the black hole is larger than the radiated luminosity, as is the jet power. They are jet dominated advective systems. 

\citet{MerloniHeinzdiMatteo2003} showed that also those AGN, that are likely to belong to the soft state, seem to follow the fundamental plane in contrast to the case of XRBs. If the fundamental plane holds for hard and soft state objects, the scaling of the X-ray emission of BHs with $(\dot{M}/\dot{M}_{Edd})^2$ (see Fig.~\ref{fiMdotRLX}, \ref{FigAGN}) has to continue up to the Eddington limit. This would only be possible if all AGN have inefficient flows, which is unlikely due to the "big blue bump" found in many systems, a feature attributed to a standard disc. 

As the scaling of the inefficient flows crosses the linear scaling of efficient accretion discs at around $10\%$ Eddington (Figs 3 - 5), we only expect a reduction of the X-ray luminosity by a factor of 10. On the other hand, many of the soft state AGN are radio-quiet (this is also true for the sample of \citealt{MerloniHeinzdiMatteo2003}), reducing the radio power by roughly a factor of 10. For the quenching of the radio flux in AGN see also \citet{MaccaroneGalloFender2003}. Thus, both effects will partly average out and only increase the scatter as shown in \citet{KoerdingFalckeCorbel2005}. One can also turn this argument around: As we do not see a strong break in the fundamental plane for soft state objects, the intersection of the inefficient flow scaling and the linear scaling of the efficient discs can not happen at $\dot{M} \leq 10^{-3}$ Eddington accretion rates. This would be detected in the fundamental plane including soft state objects. Thus, the crossing point has to be similar in AGN and XRBs and will be in the range of $2-10\%$ Eddington. This in turn suggests that hard to soft state transitions, and vice versa, will occur at about the same Eddington ratio in all BHs.

\section{Conclusions}
We have shown that the flat spectrum radio luminosity of accreting objects is a good tracer of the accretion rate. All BH XRBs with a flat spectrum radio core (a signature of the jet) seem to obey
\begin{equation}
\dot{M} = 4 \times 10^{17}  \left(\frac{L_{\mathrm Rad}}{10^{30} \mbox{erg s}^{-1}} \right)^{12/17} \left( \frac{0.1}{\eta}\frac{0.75}{f} \right) \frac{\mbox{g}}{\mbox{s}}, 
\end{equation}
while for NSs the normalisation factor is $7.7\times 10^{17}$.
This accretion measure is valid for hard state NS and BH XRBs as well as for IMS BHs with an optically thick radio jet. We have seen that this relation may also hold for AGN. This suggests that the jet launching mechanism seems to be an intrinsic feature of the accretion flow and does not depend strongly on special features of a BH (e.g., the ergo-sphere, the spin) or the NS (e.g., surface dipole, magnetic fields).

Using this accretion measure, we could show the different dependence of the bolometric luminosity on the accretion rate for BHs and NSs: While the bolometric luminosity of NSs and IMS BHs depend linearly on the accretion rate, it is roughly quadratic for hard state BHs. 

A similar picture is also found in AGN. The X-ray emission of strongly sub-Eddington systems scales quadratically with accretion rate in Eddington units. As the scaling tracks of inefficient and efficient flows intersect at accretion rates of roughly 10 $\%$ Eddington, it is not yet possible to observe the turnover from the inefficient flow to the standard discs in AGN, as the magnitude of the effect is smaller than the scatter in the correlation.

In considering also advection, we have advanced the 'jet-dominated' scenario of \citet{FenderGalloJonker2003}
The difference of the luminosities between the NSs and the BHs are likely to be advected into the black hole -- if the accretion disc winds are similar for black holes and NSs at a given Eddington ratio. This supports the idea that the sources classified as BH indeed have an event horizon. Hard state BHs are advective systems, where the jet power exceeds the X-ray luminosity at low accretion rates.

\bigskip
{\it Acknowledgements:} The authors thank Tom Maccarone and Christian Knigge for helpful discussions. We thank our referee for constructive comments.

\label{lastpage}

\bibliographystyle{mn2e}

\end{document}